\documentclass[twocolumn,showpacs,prl,amsmath,amssymb]{revtex4}

\usepackage{graphicx,bm}
\usepackage{epsfig}

\begin{document}

\title{A universal Hamiltonian for a quantum dot in the presence of spin-orbit interaction}

\author{Y.\ Alhassid and T.\ Rupp}
\affiliation{Center for Theoretical Physics, Sloane Physics
  Laboratory, Yale University,  New Haven, Connecticut 06520}

\begin{abstract}
 We derive a universal Hamiltonian for a quantum dot in the presence of spin-orbit interaction in the symmetry limits of a strong spin-dependent Aharonov-Bohm-like term. We also derive a closed expression for the conductance through such a  dot, and use it to study the effects of spin-orbit on the conductance peak statistics in the presence of an exchange interaction.  For a realistic strength of the exchange interaction, we find that the width of the peak-spacing distribution is sensitive to spin-orbit coupling only in the absence of an orbital magnetic field.   We also find that spin-orbit coupling modifies the shape of the peak-spacing distribution and suppresses the peak-height fluctuations.
\end{abstract}

\maketitle

  The statistical fluctuations of the single-particle spectrum and wavefunctions in a chaotic or diffusive quantum dot with a large Thouless conductance $g_T$ can be described by random matrix theory (RMT) \cite{alhassid00,guhr98}.  In open dots, which  are strongly coupled to leads, RMT can successfully describe the mesoscopic fluctuations of the conductance assuming non-interacting electrons. 
 However, in almost-isolated dots, which are weakly coupled to leads,  electron-electron interactions cannot be ignored. In such dots, most of the residual interaction terms are suppressed in the limit of a large $g_T$, except for a few terms. These terms constitute the interacting part of the  universal Hamiltonian \cite{kurland00,aleiner02}, and include, in addition to the charging energy, a constant exchange interaction.   Using RMT to describe the single-particle Hamiltonian, a significantly better agreement with  the data of Refs.~\onlinecite{patel98a,patel98b} is found once this exchange term is included \cite{exchange,usaj03}. In general, the exchange interaction suppresses fluctuations of both the conductance peak spacings and peak heights.

  Spin-orbit (SO) scattering is also expected to affect the conductance fluctuations, although its effects are suppressed in small dots. Enhanced SO coupling in a parallel magnetic field explained the suppression of conductance fluctuations in chaotic open 2D GaAs dots \cite{folk01,halperin01}.  The possible symmetries of the single-particle Hamiltonian in such dots with SO scattering were recently classified \cite{aleiner01}, and their signatures were observed in open dots  \cite{zumbuhl02}.  

   Realistic studies of SO effects in almost-isolated dots require the inclusion of interactions.  However, the universal  Hamiltonian for a 2D dot has been derived only in the absence of SO coupling. Here we study how the universal Hamiltonian is modified in the presence of SO scattering, both in the presence and absence of an orbital magnetic field.  In particular, we derive the universal Hamiltonian [Eq.~(\ref{universal-H})] in the new symmetry limits introduced by the leading order SO term (a spin-dependent Aharonov-Bohm-like term). We also describe the dot's Hamiltonian [Eqs.~(\ref{crossover-H}),(\ref{spin})] in the crossover induced by this SO term. Electron correlations in the presence of SO scattering were recently studied in metal nanoparticles \cite{gorokhov03}, but such 3D nanoparticles do not possess the new symmetries considered here. In addition, we derive a closed formula for the conductance through a dot described by the new universal Hamiltonian [Eqs.~(\ref{conductance}) -- (\ref{spin-prob})], and study the corresponding conductance peak statistics. For a realistic strength of the exchange interaction, we find that the standard deviation of the peak-spacing fluctuations is sensitive to SO scattering only in the absence of an orbital magnetic field. SO coupling also modifies the shape of the peak-spacing distribution and suppresses the peak height fluctuations. 

The symmetries of the one-body Hamiltonian in the presence of SO in GaAs dots were classified by applying a suitable unitary transformation to its original form \cite{aleiner01}. The transformed Hamiltonian is expanded 
 in the parameter $L/\lambda$ ($L$ is the linear size of the dot and $\lambda$ is a mean
SO scattering length), which is assumed to be small.  The two leading 
contributions are described by effective vector potentials that modify 
the vector potential of the orbital magnetic field $B$.  
The first is a spin-dependent Aharonov-Bohm-like vector potential ${\bf a}_\perp$ of order $(L/\lambda)^2$, and the second is a spin-flip term ${\bf a}_\parallel$ of order $(L/\lambda)^3$.  In small dots, only the first term is relevant and the total effective orbital field is given by
\begin{equation}\label{effective-B}
B_{\rm eff} = B + B_{\rm so} s_z\;; \;\;\; B_{\rm so}={c\hbar \over e \lambda^2}\;,
\end{equation}
where $s_z$ is the electron spin component 
 perpendicular to the plane of the dot, and $B_{\rm so}$ is an effective SO field.
The effect of the SO interaction is described by a dimensionless parameter $x^2_\perp = \kappa g_T (\Phi_{\rm so}/\Phi_0)^2=\kappa g_T ({\cal A}/\lambda^2)^2$, where ${\cal A}$ is the area of the dot, $\kappa$ is a geometrical coefficient, $\Phi_{\rm so}$ is the flux associated with the SO field, and $\Phi_0=c\hbar/e$ is the unit flux.  In the following we derive the universal Hamiltonian in the 
symmetry limits $x_\perp \gg 1$ for both $B=0$  and $B \neq 0$. 
(in practice, the crossover is often achieved for $x_\perp \sim 1$). 

 The single-particle eigenstates in the presence of an effective field (\ref{effective-B}) are 
given by $|\alpha \sigma\rangle$, where $\sigma=\pm$ describes spin up/down electrons with orbital wavefunctions 
$\psi_{\alpha \pm}({\bf r})$ and energies $\epsilon_{\alpha \pm}$. In the absence of SO coupling ($x_\perp=0$), 
$\psi_{\alpha +} =\psi_{\alpha -}$. When SO is present, we have to distinguish between two cases. For $B \neq 0$, the wavefunctions  $\psi_{\alpha \pm}({\bf r})$ correspond
to two different values 
$B \pm B_{\rm so}/2$ of the effective field, and become uncorrelated for $x_\perp \gg 1$. Thus, we have a crossover from two degenerate Gaussian unitary ensembles (GUE) at $x_\perp=0$ to two uncorrelated GUE at $x_\perp \gg 1$ \cite{orbital}.  However, for $B=0$, time reversal invariance leads to Kramers degeneracy 
$\epsilon_{\alpha +} =\epsilon_{\alpha -}$ and
$\psi_{\alpha -} (\bf r) =  \psi^*_{\alpha +}(\bf r)$, and the crossover is from two degenerate Gaussian orthogonal ensembles (GOE) at $x_\perp=0$ to two degenerate GUE at $x_\perp \gg 1$. 

 Since $|\alpha \sigma \rangle$ is a complete single-particle basis, we can write the dot's Hamiltonian in this basis.  We first discuss the limits $x_\perp=0$ and $x_\perp \gg 1$ (more precisely, $x_\perp^2 \agt g_T$)  but not the crossover itself. In these limits, we consider the ``diagonal'' part of the interaction, i.e., direct terms  $v_{\alpha \sigma \gamma \sigma';\alpha \sigma \gamma \sigma'}$, exchange terms $v_{\alpha \sigma \gamma \sigma';\gamma \sigma \alpha \sigma'}$, and Cooper channel terms  $v_{\alpha + \alpha -;\gamma + \gamma-}$. 
For $g_T \gg 1$, we separate this ``diagonal'' interaction into average and fluctuating parts, and identify terms that remain finite for $g_T \to \infty$.  We define the following average matrix elements (for $\alpha\neq \gamma$)
\begin{eqnarray}\label{averages}
v_1 &= & \bar v_{\alpha \sigma \gamma \sigma';\alpha \sigma \gamma \sigma'}\;;\;\; v_2  =  \bar v_{\alpha \sigma \gamma \sigma;\gamma \sigma \alpha \sigma} \;; \nonumber \\ 
v_3 &  = & \bar v_{\alpha \sigma \gamma -\sigma;\gamma \sigma \alpha -\sigma} \;;\;\; v_4  = \bar v_{\alpha + \alpha -;\gamma + \gamma-} \;.
\end{eqnarray}

While all the direct matrix elements have the same average $v_1$, we have distinguished two types of exchange matrix elements. For $x_\perp=0$, the orbital wavefunctions are spin-independent and $v_2=v_3=J_s$. For $x_\perp \to \infty$, $v_2=J_s$ remains unchanged (the orbital wavefunctions correspond to the same spin $\sigma$), 
but $v_3 =0$. The vanishing of $v_3$ can be shown, e.g., in a contact model for the screened interaction [$v(\bf r - \bf r') \propto \delta(\bf r - \bf r')$] for which 
$v_{\alpha \sigma \gamma -\sigma;\gamma \sigma \alpha -\sigma} \propto  \int d {\bf r}\psi^*_{\alpha \sigma}({\bf r}) \psi^*_{\gamma -\sigma}({\bf r}) \psi_{\gamma \sigma}({\bf r}) \psi_{\alpha -\sigma}({\bf r})$.  For $B \neq 0$, the correlator of the wavefunctions $\psi_{\alpha \sigma}$ and $\psi_{\alpha -\sigma}$ is a GUE parametric correlator which decays as a power law for $x_\perp \gg 1$ \cite{alhassid95,wilkinson95}. For $B=0$, the orbital wavefunctions have GUE symmetry with $\psi_{\alpha -\sigma}= \psi^*_{\alpha \sigma}$ and thus  $v_3 \propto \int d {\bf r}\overline{\psi^*_{\alpha \sigma}({\bf r}) \psi_{\gamma \sigma}({\bf r}) \psi_{\gamma \sigma}({\bf r}) \psi^*_{\alpha \sigma}({\bf r})}=0$ \cite{average}.

\begin{widetext}
Using Eqs.~(\ref{averages}),  we can write the average part of the diagonal interaction in the form
\begin{eqnarray}\label{average-int}
\bar V_{\rm diag} =   \sum_{\alpha \neq \gamma} ( \frac{1}{2} v_1 \hat n_\alpha \hat n_\gamma - \frac{1}{2} v_2 \sum_{\sigma}\hat n_{\alpha \sigma} \hat n_{\gamma \sigma} 
 + \frac{1}{2} v_3 \sum_{\sigma} a^\dagger_{\alpha \sigma} a^\dagger_{\gamma -\sigma} a_{\alpha -\sigma}a_{\gamma \sigma} + v_4 \hat T^\dagger_\alpha \hat T_\gamma) 
 - (v_1 + v_3 + v_4)\sum_\alpha \hat T^\dagger_\alpha \hat T_\alpha \;,
\end{eqnarray}
\end{widetext}
where $\hat n_{\alpha \sigma}$ is the occupation operator of the state $|\alpha \sigma\rangle$ and $\hat n_{\alpha}= \hat n_{\alpha +} +  \hat n_{\alpha -}$. Also $\hat T^\dagger_\alpha= a^\dagger_{\alpha +} a^\dagger_{\alpha -}$ is a pair creation operator, and we have used $\overline{ v_{\alpha + \alpha -;\alpha + \alpha-}} = v_1+ v_3+ v_4$ (as is easily verified for a contact interaction).  

The total spin operator of the dot can be represented in the basis $|\alpha \sigma\rangle$ as ${\bf\hat S} =\frac{1}{2} \sum_{\alpha \gamma; \sigma \sigma'} \langle \alpha \sigma | \gamma \sigma'\rangle a^\dagger_{\alpha \sigma} {\bm \sigma}_{\sigma \sigma'} a_{\gamma \sigma'}$ (${\bm \sigma}$ are Pauli matrices), and is no longer diagonal in the orbital label $\alpha$. However, since  $\langle \alpha \sigma | \gamma \sigma\rangle = \delta_{\alpha \gamma}$, the $z$ component of the total spin is diagonal $\hat S_z=\frac{1}{2}\sum_{\alpha} (\hat n_{\alpha +} - \hat n_{\alpha -}) = \sum_{\alpha} \hat s_{\alpha z}$, and we have
$\sum_\sigma \hat n_{\alpha \sigma} \hat n_{\gamma \sigma}  = \frac{1}{2} \hat n_{\alpha} \hat n_{\gamma} + 2 \hat s_{\alpha z} \hat s_{\gamma z}$. The $v_3$ term in Eq.~(\ref{average-int}) is absent in the limit $x_\perp \to \infty$, while for $x_\perp=0$ it can be related to the spin operators through $a^\dagger_{\alpha +} a^\dagger_{\gamma -} a_{\alpha -}a_{\gamma +} + a^\dagger_{\alpha -} a^\dagger_{\gamma +} a_{\alpha +}a_{\gamma -} = 
4( \hat s_{\alpha z} \hat{\bf s}_{\gamma z} - \hat s_\alpha \cdot \hat{\bf  s}_\gamma)$.  Thus in both symmetry limits $x_\perp=0$ and $x_\perp \to \infty$ we can write (\ref{average-int}) in the form
\begin{eqnarray}
\bar V_{\rm diag}= \frac{1}{2}(v_1- v_2/2) \hat n^2 & - & \frac{1}{2}(v_1-v_2) \hat n -v_3 \hat{\bf S}^2 \nonumber \\ & - &(v_2-v_3) \hat S_z^2 +v_4 \hat T^\dagger \hat T \;,
\end{eqnarray}
where $\hat T^\dagger=\sum_\alpha \hat T^\dagger_{\alpha}$. 
In the following we denote $\beta=1$ ($\beta=2$) for $B=0$ ($B \neq 0$). For $x_\perp=0$, $v_2=v_3=J_s$ and $v_4=\delta_{\beta 1} J_c$ ($J_c$ is the strength of the Cooper channel interaction), and we recover the  universal Hamiltonian \cite{kurland00,aleiner02}. However, for  $x_\perp \gg 1$, we have $v_3=0$ and the ${\bf \hat{S}}^2$ interaction is replaced by $\hat S_z^2$. Furthermore, $v_4=\delta_{\beta 1} J_c$ still holds for $x_\perp\gg 1$. For $B \neq 0$, $v_4$ is obviously zero, while for $B=0$ we use $\psi_{\alpha -}=\psi^*_{\alpha +}$ to obtain
$v_{\alpha + \alpha -;\gamma + \gamma-}\propto \int d {\bf r} |\psi_{\alpha +} ({\bf r})|^2 |\psi_{\gamma +}({\bf r})|^2$ and thus $v_4\neq 0$.

 As in the absence of SO, the off-diagonal elements of the residual interaction are suppressed at large $g_T$.  Thus, for $x_\perp \gg 1$, we obtain a new universal Hamiltonian 
\begin{equation}\label{universal-H}
\hat H= \sum_{\alpha \sigma} \epsilon_{\alpha\sigma} \hat n_{\alpha \sigma}+  \frac{1}{2}U_d \hat n^2 -J_s \hat S_z^2 + \delta_{\beta 1} J_c \hat T^\dagger \hat T  \;.
\end{equation}
For $B=0$, the spin up/spin down levels are degenerate GUE levels ($\epsilon_{\alpha +} = \epsilon_{\alpha -}$), while for $B \neq 0$, the spin up/spin down levels are uncorrelated GUE levels. The important new feature of (\ref{universal-H}) is that the exchange interaction is now given by $-J_s \hat S_z^2$ instead of $-J_s \hat{\bf S}^2$ \cite{dipolar}. 

In the crossover itself (finite $x_\perp$), the fluctuations of the off-diagonal matrix elements are enhanced and cannot be ignored \cite{brouwer02}. Instead, we use the interaction of the universal Hamiltonian in the absence of SO scattering, and add SO coupling to the single-particle Hamiltonian 
\begin{eqnarray}\label{crossover-H}
\hat H_c = \sum_{\alpha \sigma} \epsilon_{\alpha\sigma} \hat n_{\alpha \sigma}& + & \frac{1}{2}U_d \hat n^2   - J_s \hat S_z^2 + \delta_{\beta 1} J_c \hat T^\dagger \hat T  \nonumber \\ 
& - & {1\over 2}J_s 
(\hat S_+\hat S_- + \hat S_-\hat S_+) \;,
\end{eqnarray}
where we have used $\hat{\bf S}^2 = (\hat S_+\hat S_- + \hat S_-\hat S_+)/2 +\hat S_z^2$. The spin and pairing operators can be rewritten in  the SO eigenstates $|\alpha \sigma\rangle$. The spin projection  $S_z= \frac{1}{2}\sum_\alpha (\hat n_{\alpha+} - \hat n_{\alpha -})$ and the pair operator $\hat T^\dagger = \sum_\alpha a^\dagger_{\alpha +} a^\dagger_{\alpha-}$ (for $\beta=1$) remain diagonal as in the universal Hamiltonian (\ref{universal-H}). However, the spin components in the plane of the dot acquire an off-diagonal form
\begin{eqnarray}\label{spin}
\hat S_+ = \sum_{\alpha \gamma} \zeta^*_{\gamma \alpha} a^\dagger_{\alpha +} a_{\gamma -}\;;\;\; \hat S_- = \sum_{\alpha \gamma} \zeta_{\alpha \gamma} a^\dagger_{\alpha -} a_{\gamma +} \;.
\end{eqnarray}
Here $\zeta_{\alpha \gamma}$ are fluctuating quantities whose statistical properties depend on $x_\perp$. For $B\neq 0$, $\zeta_{\alpha \gamma} \equiv \langle \alpha(0) | \gamma (x_\perp)\rangle$ describe parametric overlaps in the Gaussian unitary process (GUP) between the eigenstates at  scaled parameter values $x_\perp=0$ and $x_\perp$ \cite{alhassid95,wilkinson95}. For $B=0$, $\zeta_{\alpha \gamma}$ are just the orthogonal invariants $\rho_{\alpha \gamma}$ defined by the ``real'' scalar product of the eigenstates $\alpha$ and $\gamma$ \cite{brouwer02}. In the GUP, $\overline{|\zeta_{\alpha \gamma}|^2} = x_\perp^2/[(\epsilon_\alpha -\epsilon_\gamma)^2/\Delta^2 +(\pi x_\perp^2)^2]$, where $\epsilon_\alpha$ are the energy levels at a parameter value $x_\perp \gg 1$ \cite{wilkinson95} (a similar expression holds in the GOE to GUE crossover \cite{brouwer02}). Thus the number of levels $\gamma$ coupled to a  level $\alpha$ in (\ref{spin}) is $\sim x_\perp^2$. Since (\ref{crossover-H}) describes an effective Hamiltonian of $\sim g_T$ levels around the Fermi energy, it is valid for $x_\perp^2 \ll g_T$. The matrix elements $\zeta_{\alpha \gamma}\zeta_{\mu \nu}$ of the interaction $\hat S_+\hat S_- + \hat S_-\hat S_+$ have rms values $\lesssim 1/x_\perp^2$, and dominate the corrections to (\ref{crossover-H}) which are of the order $1/g_T$ . For $x^2_\perp \agt g_T$, the effective Hamiltonian is the universal Hamiltonian (\ref{universal-H}).  The spin $S$ is no longer a good quantum number of the Hamiltonians (\ref{universal-H}) and (\ref{crossover-H}) (for $x_\perp\neq 0$), but $S_z=M$ remains a good quantum number.  

In the following we focus on the universal Hamiltonian (\ref{universal-H}) and ignore the Cooper channel term. Since $[\hat n_{\lambda \sigma},\hat S_z]=0$, the occupations ${\bf n} \equiv \{n_{\lambda \sigma}\}$ form a complete set of good quantum numbers. The corresponding eigenstates $| N {\bf n} M \rangle$  have energies $\varepsilon^{(N)}_{{\bf n} M} = \sum_{\lambda \sigma} \epsilon_{\lambda \sigma} n_{\lambda \sigma} +U_d N^2/2 - J_s M^2$, where $N=\sum_{\lambda \sigma} n_{\lambda \sigma}$ is the total number of electrons and $M=\sum_\lambda(n_{\lambda +}-  n_{\lambda -})/2$. 

 We have calculated the conductance $G$ at temperature $T$ using the rate equations approach \cite{master}. Defining a scaled conductance $g = (\hbar k T/ e^2 \bar \Gamma) G$ ($\bar \Gamma$ is an average tunneling width), we find
\begin{equation}\label{conductance}
g = \sum_{\lambda \sigma} w_{\lambda \sigma} g_{\lambda \sigma} \;.
\end{equation}
Here $g_{\lambda \sigma}\! = (2/\bar\Gamma) \Gamma_{\lambda \sigma}^{\rm l} \Gamma_{\lambda \sigma}^{\rm r}/
(\Gamma_{\lambda \sigma}^{\rm l} + \Gamma_{\lambda \sigma}^{\rm r})$ are the
single-particle level conductances, where $\Gamma^{\rm r}_{\lambda \sigma}$ ($\Gamma^{\rm r}_{\lambda \sigma}$) is the partial width of an electron in level $\lambda \sigma$ to decay to the left (right) lead. The thermal weights $w_{\lambda \sigma}$ are given by
\begin{equation}\label{weight}
w_{\lambda \sigma} = \sum_{{\bf n} \atop n_{\lambda \sigma}=0} \tilde P^{(N)}_{{\bf n} M} f(\varepsilon^\sigma_{\lambda M}) \;,
\end{equation}
where the sum over all occupation sequences is restricted to the level $\lambda$ with spin $\sigma$ being empty. In Eq. (\ref{weight}),  $f(x)=(1+e^{x/kT})^{-1}$ is the Fermi-Dirac function evaluated at  an energy $\varepsilon^\pm_{\lambda M}= \epsilon_{\lambda \pm}  \mp J_s (M \pm 1/4) -\tilde \epsilon_F$, which corresponds to the addition of an electron with spin $\sigma$ is to a level $\lambda$ (${\tilde
  \epsilon}_{\rm F}$ is an effective Fermi energy).  The quantity $\tilde P^{(N)}_{{\bf n} M}= e^{-[\varepsilon^{(N)}_{{\bf n} M} -\tilde \epsilon_F N]/kT}/Z$ is the equilibrium probability of the state $| N {\bf n} M\rangle$, with a partition function $Z$ defined as a Boltzmann-weighted
sum over all possible $N$- and $(N+1)$-body states.  The sum over all occupation numbers in Eq.~(\ref{weight}) can be evaluated in closed form at constant $M$. The constraint $n_{\lambda \sigma}=0$ is taken into account by introducing the factor $1-n_{\lambda \sigma}$, and we have
\begin{equation}\label{w-sigma}
w_{\lambda \sigma} = \sum_M  (1-\langle n_{\lambda}\rangle^\sigma_{n_{\sigma}}) \tilde P_{N, M} f(\varepsilon^\sigma_{\lambda M}) \;;
\end{equation}
where $-N/2 \leq M \leq N/2$. 
Here $n_{\pm }= N/2 \pm M$ is the number of spin-up and spin-down electrons, and $\tilde P_{N,M}$ is the probability to find the dot with $N$ electrons and spin projection $S_z=M$. Since the trace at fixed $N,M$ is equivalent to a trace at fixed $n_+,n_-$, we find
\begin{equation}\label{spin-prob}
\tilde P_{N,M} = e^{-( F^{+}_{n_+} + F^{-}_{n_-} + U_{N,M} )/kT}/ Z \;,
\end{equation} 
with $U_{N,M} = U_d N^2/2 - J_s M^2 - \tilde \epsilon_{\rm F} N$.
In Eqs.~(\ref{w-sigma}) and (\ref{spin-prob}) $F^{\sigma}_{n_\sigma} = -kT \ln {\rm tr}_{n_\sigma} e^{-
  \sum_{\lambda} \epsilon_{\lambda \sigma} c^\dag_\lambda
  c_\lambda/kT}$ and $\langle n_{\lambda}\rangle^\sigma_{n_{\sigma}}$ are  free energy and canonical occupations of $n_\sigma$ non-interacting
spinless fermions with energies $\epsilon_{\lambda \sigma}$.

\begin{figure}
\epsfxsize= 0.98 \columnwidth
\epsfbox{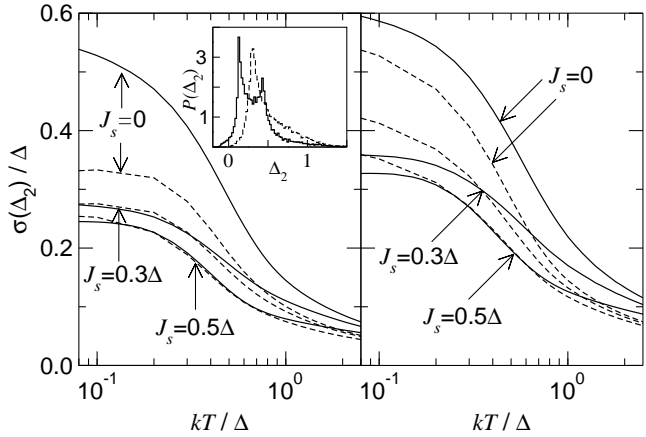}
\caption{The width $\sigma(\Delta_2)$ of the peak-spacing distribution 
(in units of $\Delta$)  for several values $J_s$ of the exchange interaction and in the presence (left panel) and absence (right panel) of an orbital magnetic field. The results in the presence of SO coupling for the universal Hamiltonian (\ref{universal-H}) (dashed lines), are compared with the results in the absence of SO coupling (solid lines). Inset: the peak-spacing distribution $P(\Delta_2)$ ($\Delta_2$ is measured relative to $e^2/C$)  in the limits $x_\perp=0$ (solid) and  $x_\perp \gg 1$ (dashed) for $J_s=0.3 \Delta$ and $k T=0.2 \Delta$.}
\label{fig1}
\end{figure}

Effects of the SO interaction on the conductance peak height statistics in the absence of exchange interaction were discussed in Ref.~\onlinecite{held03}. Here we consider SO effects on both the peak spacing and peak height statistics in the presence of exchange. We calculated these statistics in the limit $x_\perp \gg 1$ using the universal Hamiltonian (\ref{universal-H}), and compared them with the corresponding statistics in the absence of SO interaction (i.e., $x_\perp =0$) \cite{exchange}. The left panel of Fig.~\ref{fig1} shows the width $\sigma(\Delta_2)$ of the peak spacing distribution versus $k T/\Delta$ for several values $J_s$ of the exchange interaction and for both limits  $x_\perp \gg 1$ (dashed lines) and $x_\perp=0$ (solid lines).  In the absence of exchange ($J_s=0$), the SO interaction leads to a strong suppression of the spacing fluctuations. However, at $J_s=0.3 \Delta$ the width is no longer sensitive to the SO coupling, except for a small suppression at higher temperatures. 
 The peak-spacing distribution itself is affected by SO coupling and decreases more slowly at large spacings. In particular, the bimodality of the $x_\perp=0$ distribution for $kT \alt 0.3\ \Delta$ disappears for $x_\perp \gg 1$ (see inset in Fig.~\ref{fig1}) \cite{bimodal}. The right panel of Fig.~\ref{fig1} shows similar results as in the left panel but for $B=0$. In contrast to the $B \neq 0$ case, we observe that the width $\sigma(\Delta_2)$ is still sensitive to the SO coupling at $J_s=0.3\ \Delta$. SO enhances the spacing fluctuations at low temperatures but suppresses them at higher temperatures. 

\begin{figure}[t]
 \epsfxsize=0.98\columnwidth 
 \epsfbox{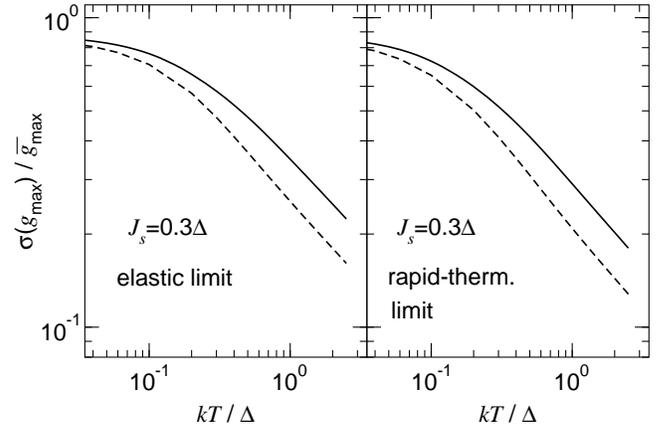}
\caption{The ratio $\sigma(g_{\rm max})/\overline{g_{\rm max}}$
  between the standard deviation and the average value of the peak
  height versus temperature $kT$ for an exchange interaction of $J_s=0.3\ \Delta$ in the presence of an orbital field and for both the elastic (left panel) and rapid-thermalization (right panel) limits.  The dashed lines correspond to the universal Hamiltonian (\ref{universal-H}), while the solid lines are the results in the absence of SO coupling \cite{exchange}.}
\label{fig2}
\end{figure}

A quantity that characterizes the peak height statistics is the ratio between the
standard deviation $\sigma(g_{\rm max})$ and the average
$\overline{g_{\rm max}}$ of the peak heights $g_{\rm max}$. The left panel of Fig.~\ref{fig2} shows this ratio as a function of $kT/\Delta$ for $J_s=0.3\ \Delta$ and $B \neq 0$ in both limits $x_\perp=0$ and $x_\perp \gg 1$. We observe that SO scattering suppresses $\sigma(g_{\rm max})/\overline{g_{\rm max}}$, and this suppression becomes stronger with temperature. 

  At higher temperatures, inelastic scattering becomes important.  The right panel of Fig.~\ref{fig2} shows similar results as in the left panel but in the rapid-thermalization limit of strong inelastic scattering. We observe that SO interaction can lead to a significant suppression of $\sigma(g_{\rm max})/\overline{g_{\rm max}}$ also in the rapid-thermalization limit. 

 At present, no statistical data are available for observing SO effects in the presence of an exchange interaction.  Using the value of $\lambda$ determined from experiments in large open dots ~\cite{zumbuhl02}, we estimate $x_\perp \ll 1$ for the small dots of Refs.~\onlinecite{patel98a,patel98b}. It would be interesting to measure the conductance peak statistics in almost-isolated dots with large area ${\cal A}$, in which SO effects are enhanced (for a fixed electron density, $x_\perp \propto {\cal A}^{5/4}$). In such large dots, it is difficult to reach the limit $T \ll \Delta$, but our results are not restricted to low temperatures. 

In conclusion, we have derived the universal Hamiltonian of a quantum dot  in the new symmetry limits when the leading order SO interaction term is included, both in the presence and absence of an orbital magnetic field. Using this universal Hamiltonian,
we have identified the signatures of SO scattering in the conductance peak statistics in the presence of an exchange interaction.
 The universal Hamiltonians for other symmetries in the presence of SO scattering and the corresponding crossover Hamiltonians will be discussed elsewhere.

We acknowledge useful discussions with B.L. Altshuler, P.W. Brouwer, V. Fal'ko, D. Huertas-Hernando, C.M. Marcus, and A.D. Mirlin. This work was supported in part by the U.S. DOE grant No.\ DE-FG-0291-ER-40608.


\begin{thebibliography}{99}

\bibitem{alhassid00}
Y.\ Alhassid, Rev.\ Mod.\ Phys.\ {\bf 72}, 895 (2000).

\bibitem{guhr98} T. Guhr, A.\ M\"uller-Groeling, and H.\ A.\ Weidenm\"{u}ller, 
 Phys. Rep. {\bf  299}, 190 (1998).

\bibitem{kurland00} 
I.~L.\ Kurland, I.~L.\ Aleiner, and B.~L.\ Altshuler, Phys.\ 
Rev.\  B {\bf 62}, 14886 (2000).

\bibitem{aleiner02} I.L. Aleiner, P.W. Brouwer, and L.I. Glazman, Phys. Rep. {\bf  358}, 309 (2002).

\bibitem{patel98a}
S.~R.\ Patel {\em et al.},
Phys.\ Rev.\ Lett.\ {\bf 80}, 4522 (1998).

\bibitem{patel98b} 
S.~R.\ Patel {\em et al.}. 
Phys.\ Rev.\ Lett.\ {\bf 81}, 5900 (1998).

\bibitem{exchange}  Y. Alhassid and T. Rupp, Phys. Rev. Lett. {\bf 91},  056801 (2003).

\bibitem{usaj03}
G.\ Usaj and H.~U.\ Baranger, Phys.\ Rev.\ B {\bf 67}, 121308 (2003).

\bibitem{folk01} J. A. Folk {\em et al.}, 
Phys. Rev. Lett. {\bf 86}, 2102 (2001).

\bibitem{halperin01} B. I. Halperin {\em et al.}, 
Phys. Rev. Lett. \textbf{86}, 2106 (2001).

\bibitem{aleiner01} I.L. Aleiner and V.I. Fal'ko, Phys. Rev. Lett.  {\bf 87}, 256801 (2001); {\bf 89}, 079902(E) (2002).

\bibitem{zumbuhl02} D. M. Zumb\"uhl {\em et al.}, 
 Phys. Rev. Lett. {\bf 89}, 276803 (2002).

\bibitem{gorokhov03} D. A. Gorokhov and P.W. Brouwer, Phys. Rev. Lett. {\bf 91}, 186602 (2003); cond-mat/0311086.

\bibitem{orbital} We assume the orbital field to be sufficiently large so that none of the effective fields in (\ref{effective-B}) vanish as $B_{\rm so}$ increases.  

\bibitem{alhassid95} Y. Alhassid and H. Attias, Phys. Rev. Lett. {\bf 74}, 4635 (1995); Phys. Rev. Lett. {\bf 76}, 1711 (1996).

\bibitem{wilkinson95} M. Wilkinson and P. Walker, J. Phys. A {\bf 28}, 6143 (1995).

\bibitem{average} The quantity $v_{\alpha \sigma \gamma -\sigma; \gamma -\sigma \alpha \sigma}$ is not gauge invariant and  its average is generally not well defined.  However in the limit $x_\perp \to\infty$, this quantity self averages to zero. In particular, $\overline{|v_{\alpha \sigma\gamma -\sigma; \gamma -\sigma \alpha \sigma}|^2}=0$ (up to $1/g_T$ corrections).

\bibitem{dipolar} The exchange interaction in (\ref{universal-H}) is formally similar to the dipolar interaction of the double dot model by  S. Adam, P. W. Brouwer, and P. Sharma, cond-mat/0309074.

\bibitem{brouwer02} S. Adam {\em et al.},
 Phys. Rev. B {\bf 66}, 165310 (2002).

\bibitem{master}
Y.\ Alhassid, T.\ Rupp, A.\ Kaminski, and L.~I.\ Glazman,
arXiv:cond-mat/0212072.

\bibitem{held03}
K.\ Held, E.\ Eisenberg, and B.~L.\ Altshuler, Phys.\ Rev.\ Lett.\
 {\bf 90}, 106802 (2003).

\bibitem{bimodal} 
Finite-$g_T$ corrections to the universal Hamiltonian can also suppress the bimodality.  

\end{thebibliography}
\end{document}